\documentclass[grl]{agu2001}
\usepackage{graphicx}
\usepackage{float}
\usepackage{subfigure}
\usepackage{amsmath}

\authorrunninghead{Robert et al.}
\titlerunninghead{Preconditioning 4D-Var data assimilation method}
\authoraddr{C. Robert,
LMC-IMAG, BP 53, 38041 Grenoble Cedex 9, France
(Celine.Robert@imag.fr)}

\begin{document}
\title{Reduced-order 4D-Var: a preconditioner for the Incremental 4D-Var data assimilation method}
\author{C. Robert, E. Blayo}
\affil{LMC-IMAG, CNRS-INPG-INRIA-UJF}
\author{J. Verron}
\affil{LEGI, CNRS-INPG-UJF}
\begin{abstract}
This study demonstrates how the incremental 4D-Var data assimilation method can be applied efficiently preconditioned in an application to an oceanographic problem. 
The approach consists in performing a few iterations of the reduced-order 4D-Var prior to the incremental 4D-Var in the full space in order to achieve faster convergence.
An application performed in the tropical Pacific Ocean, with assimilation of TAO temperature data, 
shows the method to be both feasible and efficient. It allows the global cost of the assimilation to be reduced by a factor of 2 without affecting the quality of the solution.
\end{abstract}
\begin{article}

\section{Introduction}
\label{Sect:Intro}
Computational requirements remain  a major limiting factor for operational forecasting of
atmospheric and oceanic circulations.  In such systems, most of the
computation resources are generally devoted to data assimilation: typically, sequential data 
assimilation may cost one order of magnitude more than a model simulation, and variational data assimilation two orders of magnitude more. 
Even allowing for the evolution of computer technology, such constraints seem likely to remain  for many years to come since numerous scientific studies have shown that an increase in model resolutions is needed per se. Within this context, the aim of this letter is to report on how a reduced-order approach to
variational data assimilation can help decrease the computational cost of this method.
\section{Full space and reduced order Incremental 4D-Var}
\label{Sect:4DVar}
The usual method employed for variational assimilation in current meteorological and oceanographical applications  is the incremental 4D-Var [Courtier
\textit{et al.}, 1994].
This method aims at determining an optimal correction $\delta
\textbf{x}$, defined in the full space, to a first approximation $\textbf{x}^b$ of the initial condition, which 
minimizes a functional 
\begin{eqnarray}
J(\delta \textbf{x})&=&J_b(\delta \textbf{x}) + J_o(\delta \textbf{x})\\
&=& \frac{1}{2} (\delta \textbf{x})^T \textbf{B}^{-1}\delta \textbf{x} \nonumber \\
&+ \frac{1}{2}&\sum_{i=1}^{N}({\bf{H_{i}M_i}}\delta \textbf{x} 
-\bf{d}_i)^{T} R_{i}^{-1}({\bf{H_{i}M_i}}\delta \textbf{x} -\bf{d}_i)
\label{Eq:jtot}
\end{eqnarray}
where $\textbf{M}_i$ is the tangent linear model between the initial time $t_0$ and time
$t_i$, and
$\textbf{H}_i$ is the observation operator, linearized at time $t_i$. The innovation vector, $\textbf{d}_i$, is the difference between the observation vector
$\textbf{y}_i$ and its model equivalent at time $t_i$ : $\bf{d}_i = \textbf{y}_i -
{\bf{H_iM_i\textbf{x}_b}}$.\\
The background error covariance matrix \textbf{B} is generally rather poorly known, and its
definition presents a difficult problem. This lack of suitable definition impacts strongly on the quality of the
solution and has been the subject of numerous research studies. In the usual 
incremental 4D-Var approach, \textbf{B} may be defined analytically, using, for example,
monovariate gaussian-like covariances, and balance equations for multivariate
covariances. 

In a reduced-order approach (hereafter denoted R-4D-Var), \textbf{B} and
$\delta
\textbf{x}$ are specified in a low-dimension subspace which contains a large part of the
natural variability of the system. An
efficient way to build such a subspace is to define it as the span of a few EOFs,
$\textbf{L}_i$ ($i=1,\ldots,r$), computed from previous model simulations [Blayo \textit{et al.}, 1998; Durbiano, 2001;
Robert \textit{et al.}, 2005]. 

Formally, $J_o$
remains unchanged, and only the expressions for
\textbf{$\delta$x} and \textbf{B} change in $J_b$. The increment \textbf{$\delta$x} is expanded in the EOFs
basis $  \displaystyle{\delta \textbf{x} = \displaystyle \sum_{i=1}^{r} w_i \textbf{L}_i = \textbf{Lw}}$, 
and \textbf{B} is then naturally represented by the low-rank matrix $\displaystyle{\textbf{B}_r=\textbf{L}\boldsymbol{\Lambda}_r\textbf{L}^T}$ 
with $\boldsymbol{\Lambda}_r = diag(\lambda_1, ...,\lambda_r)$ where $\lambda_i$ is the eigenvalue corresponding to $\textbf{L}_i$ 
(see Robert \textit{et al.}, 2005 for details). Thus, $\textbf{B}_r$ naturally contains 3-D
multivariate covariances. Note that another stategy consists in reducing the model itself using a POD approach [Cao \textit{et al.}, 2005; Daescu
and Navon, 2006].
\vspace*{5mm}\par
Both full space and reduced-order 4D-Var methods have been applied in the context of data
assimilation in a primitive equation model of the tropical Pacific Ocean. The model is the
OPA code [Madec \textit{et al.}, 1998] in its so-called TDH configuration, with the
variational data assimilation package OPAVAR [Weaver \textit{et al.}, 2003]. Numerous earlier studies 
have been conducted using this configuration with the incremental 4D-Var, assimilating TAO and XBT
temperature profiles, and have produced good results
[Weaver \textit{et al.}, 2003; Vialard \textit{et al.},
2003]. In these studies, the background error covariance matrix \textbf{B} is defined as
an operator including gaussian-like covariance functions, but it remains monovariate 
 [Weaver and Courtier, 2001]. Efforts to develop a multivariate 
 operator have been made recently and implemented in the 3D-Var context [Ricci \textit{et al.}, 2005].\par 
This incremental 4D-Var approach needs  quite a large number of iterations -typically 40- to
converge [Vialard \textit{et al.}, 2003, Weaver \textit{et al.}, 2003]. Since each iteration requires one run of both the
tangent linear model and the adjoint model, the computational cost can thus become
prohibitive for real configurations. In a non-linear situation such as the mid-latitude ocean, the cost will be even greater [Blum \textit{et al.}, 1998].\par
By working only in a low-dimension space, and thereby optimizing a very limited number of
coefficients (we have retained 30 EOFs vectors, explaining 92$\%$ of the total variability), the R-4D-Var tries 
in particular to overcome this drawback.
This method has proved to work well in the idealized context of twin experiments,
i.e. when the model assimilates simulated observations and is thus supposed to be perfect [Robert \textit{et al.},
2005]. The computational cost is decreased in this case by a factor of at least 3, and the
optimal increment is very well identified in the space spanned by the EOFs.
\section{A Two-Step 4D-Var strategy}
When dealing with real data, the definition of a relevant EOF basis,
representative of the true variability of the system, is much more challenging.  Since
the model has significant errors, EOFs computed from model simulations without assimilation do
not contain the right information and lead to poor R-4D-Var results. One possible way to
address this difficulty is to compute EOFs from a previous assimilated run, with another
assimilation method for example. However, since the system providing the EOFs remains
in any case imperfect, the low-dimension correction subspace does not contain all the
relevant variability of the true dynamics.
The key idea of this study is therefore to
combine the quality of the identification performed by 4D-Var and the
efficiency of R-4D-Var. This means that we continue to look for an optimal correction
$\delta \textbf{x}$ in the full space, but address the problem of the associated high
computational cost by using R-4D-Var to provide a relevant initial guess for full
space minimization. The approach consists thus in performing a few iterations of R-4D-Var
(without trying to reach a converged solution), and in using this current estimate of
$\delta \textbf{x}$ to initialize additional iterations of 4D-Var.
Only a few of the first iterations of R-4D-Var appear to be effective in providing a
relevant first guess, defined in the reduced space. This first guess can then be corrected and
improved in the full space by a few iterations of 4D-Var. This technique can thus be seen
as a  preconditioning  of 4D-Var by R-4D-Var, and will hereafter be referred to as  ``Two-Step
4D-Var" (TS-4D-Var).
Note that the
 expression of $\textbf{B}$ changes between R-4D-Var and 4D-Var. This means that the role 
of $J_b$ is identical in both methods (to minimize $\| \delta \textbf{x} \|$) but with different norms. 
The other functional $J_0$ remains unchanged. The cost functions being quadratic (due to the incremental approach), any unconstrained minimization
algorithm may be used. We use here a BFGS-like algorithm.
\section{Assessing the Two-Step strategy}
\begin{figure}[!h]
\centering
\includegraphics[width=0.475\textwidth]{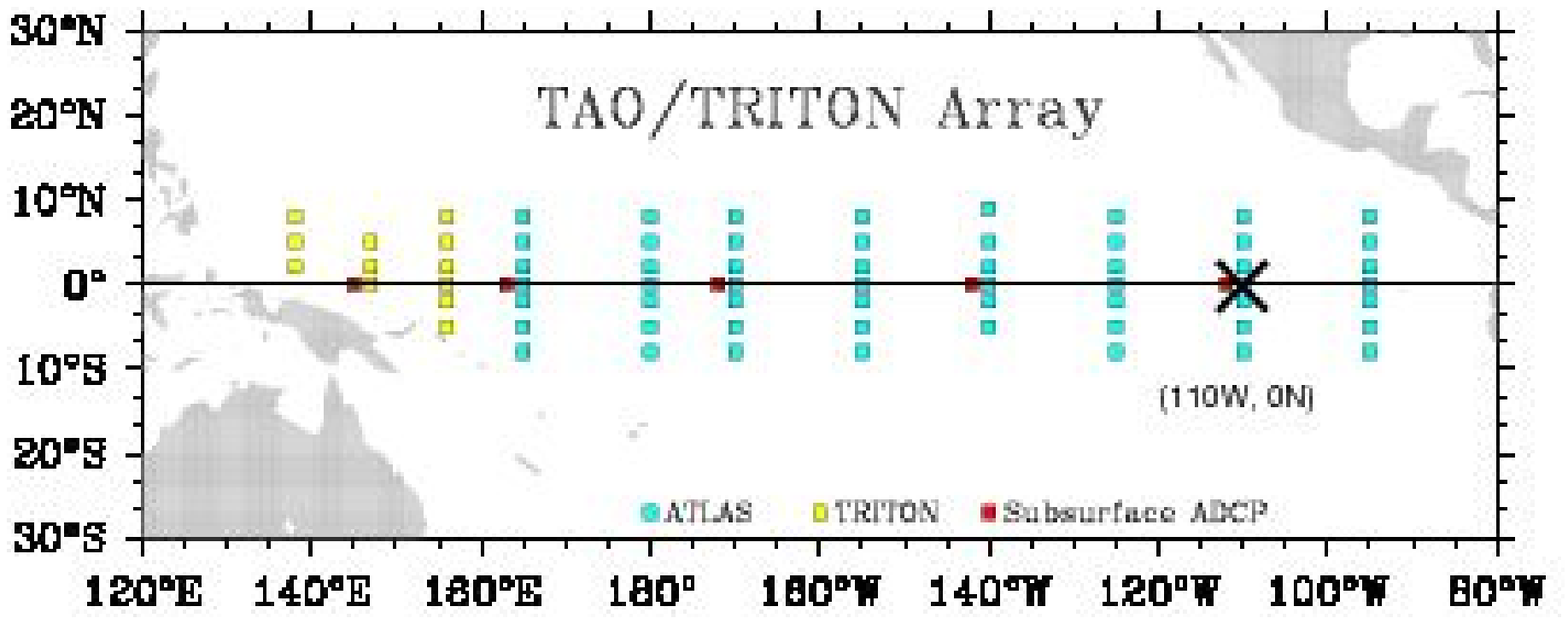}
\caption{Geographical extension of the model, location of the TAO/TRITON array points and location of the point 
chosen for the time-depth diagrams displayed in Fig. \ref{Fig:TAO-TN} and Fig. \ref{Fig:TAO-UN} (X symbol).}
\label{Fig:TDH}
\end{figure}

In order to validate this TS strategy, experiments are performed in the tropical Pacific Ocean 
(Fig. \ref{Fig:TDH}), using the OPA model in its TDH configuration. The atmospheric forcings are daily ERS-TAO winds [Menkes \textit{et al.}, 1998] 
and monthly ECMWF heat fluxes. Our experiments start in January 1993 and last one year. 
The dynamics during this period is representative of a "normal" year in the tropical Pacific, without the noteworthy influence of any ENSO
event. These conditions are seen as favorable to test the
method.\par
In order to be able to compare our results with those of previous studies performed with the same configuration [Vialard
\textit{et al.}, 2003], we assimilate temperature  data from the TAO/TRITON array plus XBT. \par
In the TS-4D-Var, we first perform 10 iterations of the R-4D-Var. These are followed 
by a run of the fully non-linear direct model in order to update the reference trajectory
for linearization. Then, 10 iterations of the incremental 4D-Var are performed. Thus, the cycle of the TS-4D-Var is like a cycle of the 4D-Var with 2 outer
loops consisting of 10 inner loops each.\par
\begin{figure}[!h]
\centering
\includegraphics[width=7cm, height=7cm, angle=-0]{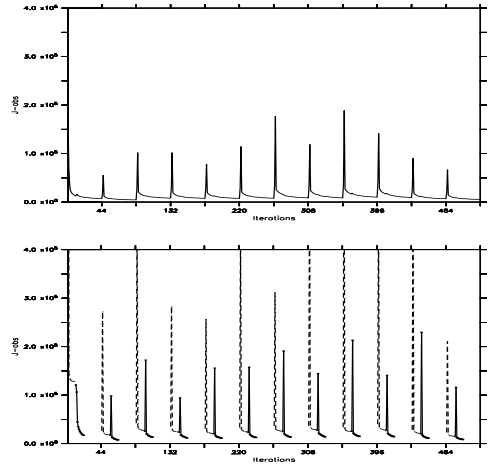}
\caption{Cost function for the observations $J_o$ for each method, as a function of time (in
 months). Top panel: 4D-Var with 44 iterations for each 
assimilation window. Bottom panel: TS-4D-Var with 22 iterations for each assimilation window 
(dashed black line: R-4D-Var, black line: Full 4D-Var).}
\label{Fig:Cost-jo}
\end{figure}

Figure \ref{Fig:Cost-jo} shows the two successive sequences of the observation term $J_o$ in the cost function.  
We can see that the R-4D-Var performs a
first descent (10 iterations, although 5-6 may be sufficient) followed by a second one performed by the incremental 4D-Var. The second
sequence allows us to reach a lower level very quickly and to stabilize the value of the cost
function after very few iterations. Thus, the minimization phase requires a lower number of 
iterations to reach the minimum (less than 20, whereas 40 are required with the incremental 4D-Var), due to the fact that we retain only the 30 largest
EOF vectors.
\par
\begin{figure}[!h]
\centering
\includegraphics[width=7cm, height=15cm]{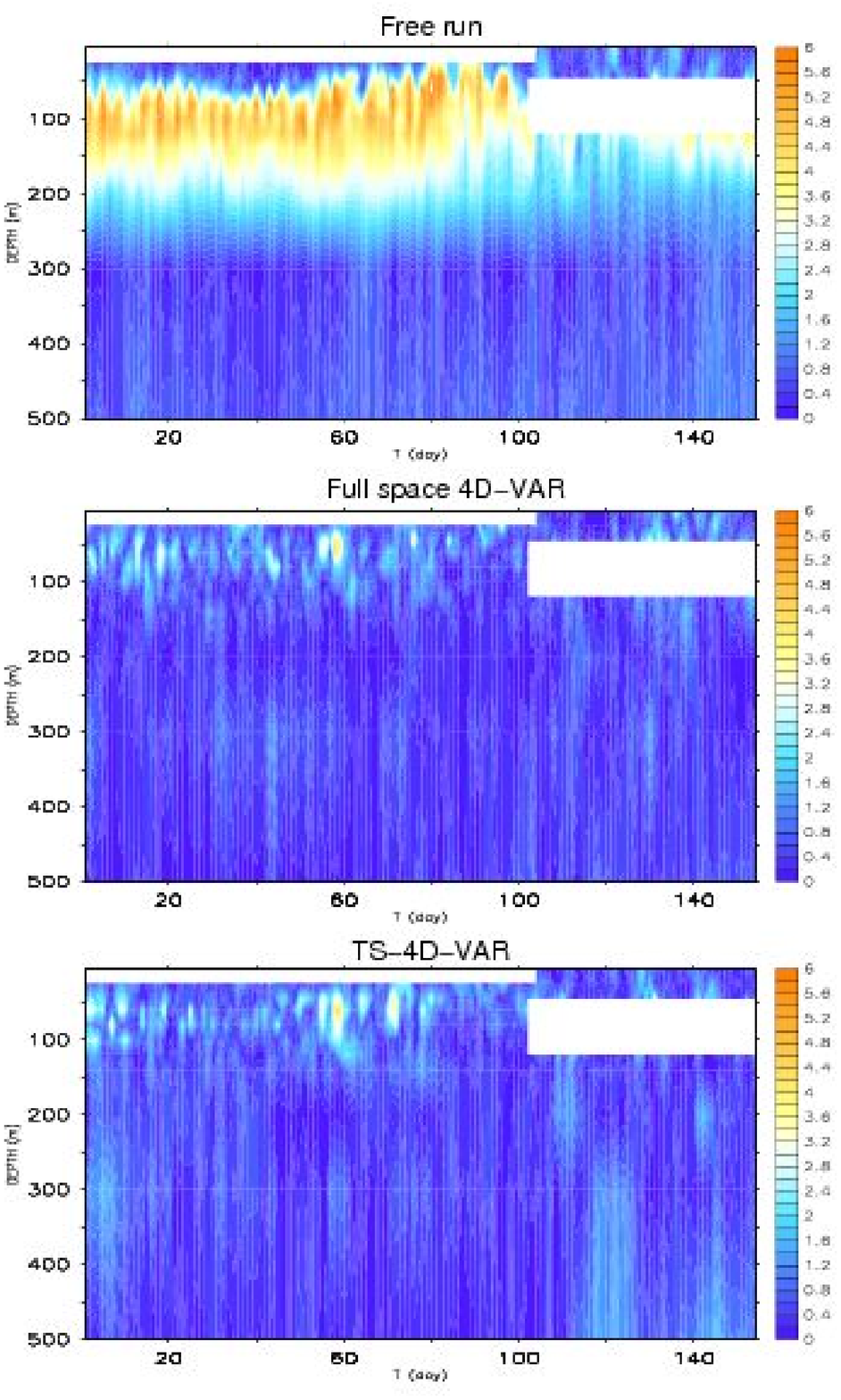}
\caption{Absolute difference in the temperature fields at ($110^oW$, $0^oN$), between model and TAO data, as a function of time (horizontal axis) and
depth (vertical axis). Top panel: free run. Middle panel: incremental 4D-Var. Bottom panel: TS-4D-Var.}
\label{Fig:TAO-TN}
\end{figure}
\begin{figure}[!h]
\centering
\includegraphics[width=7cm, height=15cm]{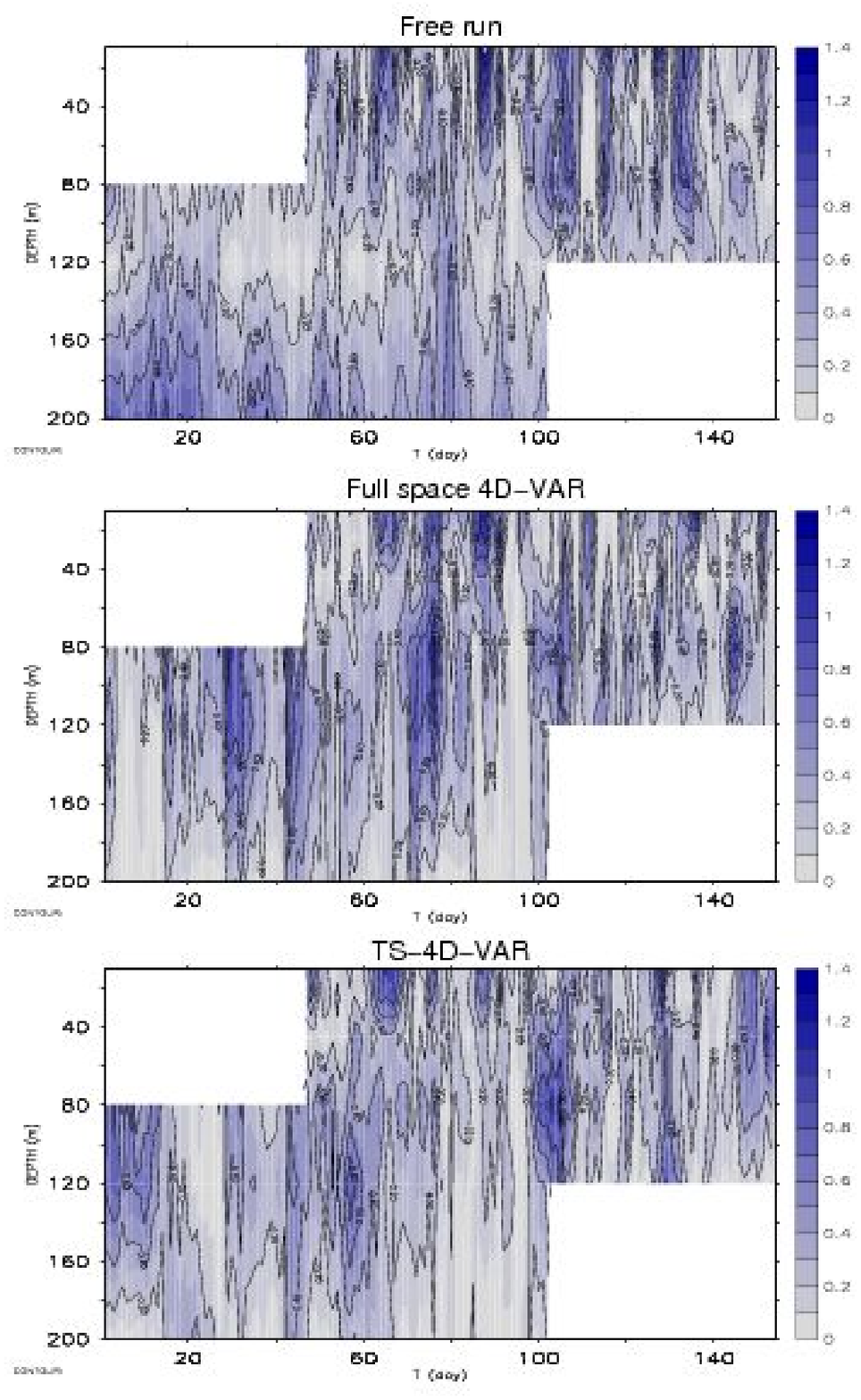}
\caption{Absolute difference in the  zonal velocity fields at ($110^oW$, $0^oN$),  between model and TAO data, as a function of time (horizontal axis) and
depth (vertical axis). Top panel: free run. Middle panel: incremental 4D-Var. Bottom panel: TS-4D-Var.}
\label{Fig:TAO-UN}
\end{figure}

For both assimilated and non-assimilated variables, the TS-4D-Var provides results that are definitely comparable to those obtained with the
incremental 4D-Var. To illustrate this point,  Figures \ref{Fig:TAO-TN} and
\ref{Fig:TAO-UN} show a comparison of the fields at a particular location in the eastern part of the domain
 at (110W, 0N)(see Fig \ref{Fig:TDH}). This location corresponds to the area where the most intense non-linearities occur. These are due to the Tropical Instability Waves which rise and propagate there, becoming increasingly intense 
from mid-June/early July. This area is thus also where identification of the
solution is the most difficult. 
 To make successive comparisons of
 the free run, the incremental 4D-Var and the TS-4D-Var with TAO data, we have plotted time-depth diagrams of the absolute 
 difference between model and TAO temperature and zonal velocity.   With regard to  temperature (Fig. \ref{Fig:TAO-TN}), the free run shows significant departures from TAO data.
The thermal field is correctly represented, however, by the two assimilation methods, which
leads to very comparable results with the same low level of absolute error, located only in the first hundred meters.
Concerning zonal velocity, which is a non-assimilated variable, the results are also satisfactory. Both assimilation methods succeed in
representing the inversion of the surface current, and the global level 
of the absolute error is of the same order of magnitude for the TS-4D-Var as for the 4D-Var, and even slightly lower (See Fig. \ref{Fig:TAO-UN}). 
\section{Conclusion}
\label{Sect:conclu}
In this letter we have proposed a new method to improve the efficiency of the 4D-Var assimilation method. Our Two-Step 4D-Var can be seen 
as a preconditioned
4D-Var, which greatly decreases the computational cost of assimilation.
This approach is validated by assimilating  real in-situ temperature profiles in a realistic model of the tropical Pacific Ocean.
The  results provided by the TS-4D-Var, for both  assimilated and non-assimilated
variables,  are of similar quality to those obtained by the 4D-Var, but the cost is divided by a factor of 2. 
For expensive configurations, this method would thus seem to be an attractive alternative to the full space 4D-Var approach.
\begin{acknowledgments}
A. Weaver provided the OPAVAR package and helped us in using it. The authors wish to thank the two reviewers whose constructive comments led to an
improvment of this paper. This work has been
supported by the CNES, the MERCATOR and the MERSEA projects. MOISE is a joint
CNRS-INPG-INRIA-UJF research project.
\end{acknowledgments}
\end{article}

\newpage
{}

\newpage

\newpage

\newpage

\newpage

\begin{thebibliography}{}

\bibitem[]{}
Blayo, E., J. Blum, and J. Verron (1998): Assimilation variationnelle de donn\'ees en oc\'eanographie et
 r\'eduction de la dimension de l'espace de contr\^ole. \textit{in Equations aux D\'eriv\'ees Partielles et 
 Applications (articles d\'edi\'es \`a Jacques-Louis Lions), Gauthier-Villars}, 199-219.
%
\bibitem[]{}
Blum, J., B. Luong and J. Verron (1998): Variational assimilation of altimeter data into a Non-linear ocean model: temporal
strategies.
\textit{ESAIM proceedings, Vol. 4, 1998, 21-57, Contrôle et Equations aux Dérivées Partielles.}
%
\bibitem[]{}
Cao, Y., J. Zhu, I.M. Navon and Z. Luo (2005): A reduced order approach to four-dimensional variational data assimilation using proper orthogonal decomposition. \textit{Submitted for publication to International Journal for Numerical Methods in Fluids.}
%
\bibitem[]{}
Courtier, P., J.N. Th\'epaut and A. Hollingsworth (1994): A strategy for operational implementation of 
4D-Var using an incremental approach. \textit{Q. J. R. Meteorol. Soc.,120}, 1367-1387.  
%
\bibitem[]{}
Daescu, D. N., I.M. Navon (2006): Efficiency of a Pod-based reduced second order adjoint model in 4D-VAR data assimilation. \textit{Submitted for publication to International Journal for Numerical Methods in Fluids.}
%
\bibitem[]{}
Durbiano S. (2001): Vecteurs caract\'eristiques de mod\`eles oc\'eaniques pour la r\'eduction d'ordre 
en assimilation de donn\'ees. \textit{PhD thesis, LMC, Joseph Fourier University of Grenoble.}
%
\bibitem[]{}
Madec, G., P. Delecluse, M. Imbard and C. Levy (1998): OPA8.1 Ocean General Circulation Model 
reference manual. \textit{IPSL Technical report.}
%
\bibitem[]{}
Menkes, C., J.P. Boulanger, A. Busalacchi, J. Vialard, P. Delecluse, 
M. J. McPhaden, E. Hackert, and N. Grima (1998): Impact of TAO vs. ERS wind stresses onto simulations of 
the tropical Pacific Ocean during the 1993-1998 period by the OPA OGCM. Climate Impact of Scale Interaction 
for the Tropical Ocean-Atmosphere System, \textit{Euroclivar Workshop Report}, \textit{13}, 46-48.
%
\bibitem[]{}
Ricci, S., A.T. Weaver, J. Vialard and P. Rogel, (2005): Incorporating state-dependent temperature-salinity constraints in the background-error
covariance of variational ocean data assimilation. \textit{Monthly Weather Review, 133}, 317-338. 
%
\bibitem[]{}
Robert, C., S. Durbiano, E. Blayo,J. Verron, J. Blum and F.-X. Le Dimet (2005): A reduced-order strategy for 4D-Var data
assimilation. 
\textit{Jour. Mar. Systems, 57(1-2)}, 70-82.
%
\bibitem[]{}		  
Vialard, J., A. Weaver, D.L.T. Anderson and
                  P. Delecluse, (2003): Three- and four-dimensional variational assimilation
                  with a general circulation model of the tropical
                  {P}acific {O}cean, Part 2: physical validation. \textit{Monthly Weather Review, 131, 1379-1395}.
%
\bibitem[]{}
Weaver, A. and P. Courtier, (2001): Correlation modelling on the sphere using a
                  generalized diffusion equation. 
                  \textit{Q. J. R. Meteorol. Soc., 127, 1815-1846}.
%
\bibitem[]{}
Weaver, A.T., J. Vialard and D.L.T. Anderson, (2003): Three- and four-dimensional variational assimilation
                  with a general circulation model of the Tropical
                  Pacific Ocean, Part 1: formulation, internal diagnostics and consistency checks. 
                 \textit{Monthly Weather Review, 131}, 1360-1378.
%
\end{thebibliography}
\end{document}